\documentclass[prd,twocolumn,showpacs,superscriptaddress]{revtex4-1}
\usepackage{graphicx}
\usepackage{units}
\usepackage{multirow}
\usepackage{bigstrut}
\usepackage{overpic}
\usepackage{array}
\usepackage{color}
\usepackage{amsmath}
\usepackage{bm}
\usepackage{times}
\RequirePackage{lineno}
\newcommand{\gevc}{\,\unit{GeV}/c}

\begin{document}

\modulolinenumbers[2]

\setlength{\oddsidemargin}{-0.5cm} \addtolength{\topmargin}{15mm}

\title{\boldmath Study of the decay $D^0\rightarrow \bar{K}^0\pi^-e^+\nu_e$}

\author{
  \small
      M.~Ablikim$^{1}$, M.~N.~Achasov$^{10,d}$, S. ~Ahmed$^{15}$, M.~Albrecht$^{4}$, M.~Alekseev$^{56A,56C}$, A.~Amoroso$^{56A,56C}$, F.~F.~An$^{1}$, Q.~An$^{53,42}$, Y.~Bai$^{41}$, O.~Bakina$^{27}$
% ...
      \\
      \vspace{0.2cm}
      (BESIII Collaboration)\\
      \vspace{0.2cm} {\it
      $^{1}$ Institute of High Energy Physics, Beijing 100049, People's Republic of China\\
% ...
$^{4}$ Bochum Ruhr-University, D-44780 Bochum, Germany\\
% ...
$^{10}$ G.I. Budker Institute of Nuclear Physics SB RAS (BINP), Novosibirsk 630090, Russia\\
% ...
$^{15}$ Helmholtz Institute Mainz, Johann-Joachim-Becher-Weg 45, D-55099 Mainz, Germany\\
% ...
$^{27}$ Joint Institute for Nuclear Research, 141980 Dubna, Moscow region, Russia\\
% ...
$^{41}$ Southeast University, Nanjing 211100, People's Republic of China\\
      $^{42}$ State Key Laboratory of Particle Detection and Electronics, Beijing 100049, Hefei 230026, People's Republic of China\\
% ...
$^{53}$ University of Science and Technology of China, Hefei 230026, People's Republic of China\\
% ...
$^{56}$ (A)University of Turin, I-10125, Turin, Italy; (B)University of Eastern Piedmont, I-15121, Alessandria, Italy; (C)INFN, I-10125, Turin, Italy\\
% ...
$^{d}$ Also at the Novosibirsk State University, Novosibirsk, 630090, Russia\\
% ...
      \vspace{0.4cm}
}
}

\begin{abstract}
We report a study of the decay $D^0 \rightarrow \bar{K}^0\pi^-e^+\nu_{e}$ based on a sample of $2.93~\mathrm{fb}^{-1}$ $e^+e^-$ annihilation data  collected at the center-of-mass energy of 3.773~GeV with the BESIII detector at the BEPCII collider. The total branching fraction is determined to be $\mathcal{B}(D^0\rightarrow \bar{K}^0\pi^-e^+\nu_{e})=(1.434\pm0.029({\rm stat.})\pm0.032({\rm syst.}))\%$, which is the most precise to date. According to a detailed analysis of the involved dynamics, we find this decay is dominated with the $K^{*}(892)^-$ contribution and present an improved measurement of its branching fraction to be $\mathcal{B}(D^0\rightarrow K^{*}(892)^-e^+\nu_e)=(2.033\pm0.046({\rm stat.})\pm0.047({\rm syst.}))\%$.
We further access their hadronic form-factor ratios for the first time as $r_{V}=V(0)/A_1(0)=1.46\pm0.07({\rm stat.})\pm0.02({\rm syst.})$ and $r_{2}=A_2(0)/A_1(0)=0.67\pm0.06({\rm stat.})\pm0.01({\rm syst.})$.
In addition, we observe a significant $\bar{K}^0\pi^-$ $S$-wave component accounting for $(5.51\pm0.97({\rm stat.})\pm0.62({\rm syst.}))\%$ of the total decay rate.
\end{abstract}

\pacs{13.30.Ce, 14.40.Lb, 14.65.Dw}

\maketitle

\section{Introduction}
The studies on semileptonic\,(SL) decay modes of charm mesons provide valuable information on the weak and strong interactions in mesons composed of heavy quarks~\cite{physrept494}. The semileptonic partial decay width is related to the product of the hadronic form factor describing the strong-interaction in the initial and final hadrons, and the Cabibbo-Kobayashi-Maskawa (CKM) matrix elements $|V_{cs}|$ and $|V_{cd}|$, which parametrize the mixing between the quark flavors in the weak interaction~\cite{prl10_531}. The couplings $|V_{cs}|$ and $|V_{cd}|$ are tightly constrained by the unitarity of the CKM matrix. Thus, detailed studies of the dynamics of the SL decays allow measurements of the hadronic form factors, which are important for calibrating the theoretical calculations of the involved strong interaction.

The relative simplicity of theoretical description of the SL decay $D\rightarrow \bar{K}\pi e^+ \nu_e$~\cite{chargeneutral} makes it a optimal place to study the $\bar{K}\pi$ system, and to further determine the hadronic transition form factors.
Measurements of $\bar{K}\pi$ resonant and non-resonant amplitudes in the decay $D^+\rightarrow K^-\pi^+e^+\nu_e$ have been reported by
the CLEO~\cite{prd74_052001}, BABAR~\cite{prd83_072001} and BESIII~\cite{prd94_032001} collaborations. In these studies a nontrival $S$-wave component is observed along with the dominant $P$-wave one.
A study of the dynamics in the isospin-symmetric mode $D^0\rightarrow \bar{K}^0\pi^-e^+\nu_e$ will provide complementary information on the $\bar{K}\pi$ system. Furthermore, the form factors in the $D\rightarrow Ve^+\nu_{e}$ transition, where $V$ refers to a vector meson, have
been measured in decays of $D^+\rightarrow \bar{K}^{*0}e^+\nu_e$~\cite{prd74_052001,prd83_072001,prd94_032001}, $D\rightarrow \rho e^+\nu_e$~\cite{prl110_131802}
and $D^+\rightarrow \omega e^+\nu_e$~\cite{prd92_071101}, while no form factor  in $D^0\rightarrow K^{*}(892)^-e^+\nu_e$ has been studied yet.
Therefore, the study of the dynamics in the decay $D^0\rightarrow K^{*}(892)^- e^+\nu_e$ provides essentially additional information on the family of $D\rightarrow V e^+\nu_e$ decays.

In this paper, an improved measurement of the absolute branching fraction~(BF) and the first measurement of the form factors of the decay $D^0\rightarrow \bar{K}^0\pi^-e^+\nu_e$ are reported.
These measurements are performed using an $e^+e^-$ annihilation data sample corresponding to an integrated luminosity of
$2.93~\mathrm{fb}^{-1}$ produced at $\sqrt{s}=3.773$ GeV with the BEPCII collider and collected with the BESIII detector~\cite{Ablikim:2009aa}. 

\section{BESIII Detector and Monte Carlo Simulation}
The BESIII detector is a cylindrical detector with
a solid-angle coverage of 93\% of $4\pi$. The detector consists of a Helium-gas based main drift chamber (MDC), a plastic
scintillator time-of-flight (TOF) system, a CsI(Tl) electromagnetic
calorimeter (EMC), a superconducting solenoid providing a 1.0\,T
magnetic field and a muon counter. The charged particle momentum
resolution is 0.5\% at a transverse momentum of 1\,$\gevc$. The
photon energy resolution in EMC is 2.5\% in the barrel and 5.0\% in the end-caps at energies of 1\,GeV.
More details about the design and performance of the detector are given in
Ref.~\cite{Ablikim:2009aa}.

A {\sc geant4}-based~\cite{geant4} simulation package,
which includes the geometric description of the detector and the
detector response, is used to determine signal detection efficiencies and
to estimate potential backgrounds. The production of the $\psi(3770)$,
initial state radiation production of the $\psi(2S)$ and $J/\psi$, and the continuum processes $e^+e^-\rightarrow \tau^+\tau^-$ and $e^+e^-\rightarrow q\bar{q}$ ($q=u$,~$d$ and $s$) are simulated with the event
generator {\sc kkmc}~\cite{kkmc}. The known decay modes are generated
by {\sc evtgen}~\cite{nima462_152} with the branching fractions set to the world-average values from the Particle Data Group~\cite{pdg16},
while the remaining unknown decay modes are modeled by
{\sc lundcharm}~\cite{lundcharm}. 
The generation of simulated signals $D^0\rightarrow \bar{K}^0\pi^-e^+\nu_e$ incorporates knowledge of the form factors, which are obtained in this work.

\section{Analysis}
The analysis makes use of both ``single-tag'' (ST) and ``double-tag'' (DT) samples of $D$ decays. The single-tag sample is reconstructed in one of the final states listed in Table~\ref{tab:numST}, which are called the tag decay modes. Within each ST sample, a subset of events is selected where the other tracks in the event are consistent with the decay $D^0\rightarrow \bar{K}^0\pi^-e^+\nu_e$. This subset is referred as the DT sample. For a specific tag mode $i$, the ST and DT
event yields are expressed as
$$N^{i}_{\rm ST}=2N_{D^0\bar{D}^0}\mathcal{B}^i_{\rm ST}\epsilon^i_{\rm ST},
~~~
N^{i}_{\rm DT}=2N_{D^0\bar{D}^0}\mathcal{B}^i_{\rm ST}\mathcal{B}_{\rm SL}\epsilon^i_{\rm DT},$$
where $N_{D^0\bar{D}^0}$ is the number of $D^0\bar{D}^0$ pairs, $\mathcal{B}^i_{\rm ST}$ and
$\mathcal{B}_{\rm SL}$ are the BFs of the $\bar{D}^0$ tag decay mode $i$ and the $D^0$ SL decay
mode, $\epsilon^i_{\rm ST}$ is the efficiency for finding the tag candidate, and
$\epsilon^i_{\rm DT}$ is the efficiency for simultaneously finding the tag $\bar{D}^0$ and the SL decay.
The BF for the SL decay is given by
\begin{equation}
\mathcal{B}_{\rm SL}=\frac{N_{\rm DT}}{\sum_i N^{i}_{\rm
ST}\times\epsilon^i_{\rm DT}/\epsilon^i_{\rm ST}}=\frac{N_{\rm DT}}{N_{\rm ST}\times\epsilon_{\rm SL}}, \label{eq:branch}
\end{equation}
where $N_{\rm DT}$ is the total yield of DT events, $N_{\rm ST}$ is the total ST yield, and
$\epsilon_{\rm SL}=(\sum_i N^{i}_{\rm ST}\times\epsilon^i_{\rm DT}/\epsilon^i_{\rm ST})/\sum_i N^{i}_{\rm ST}$
is the average efficiency of reconstructing the SL decay, weighted by the measured yields of tag modes in data.

Selection criteria for photons, charged pions and charged kaons are the same as those used in
Ref.~\cite{prl118_112001}. To reconstruct a $\pi^0$ candidate in the decay mode $\pi^0\rightarrow \gamma\gamma$, the invariant mass of the
candidate photon pair must be within $(0.115,~0.150)$~GeV$/c^2$. To improve the momentum resolution, a kinematic fit is performed to constrain the $\gamma\gamma$ invariant mass to the
nominal $\pi^0$ mass~\cite{pdg16}. The $\chi^2$ of this kinematic fit is required to be less than 20. The
fitted $\pi^0$ momentum is used for reconstruction of the $\bar{D}^0$ tag candidates.

The ST $\bar{D}^0$ decays are identified using the beam
constrained mass,
\begin{equation}
M_{\rm BC}=\sqrt{(\sqrt{s}/2)^2-|\vec {p}_{\bar D^0}|^2},
\end{equation}
where $\vec {p}_{\bar D^0}$ is the momentum of the
$\bar{D}^0$ candidate in the rest frame of the initial $e^+e^-$ system. 
To improve the purity of the tag decays, the
energy difference $\Delta E=\sqrt{s}/2-E_{\bar{D}^0}$ for
each candidate is required to be within approximately
$\pm3\sigma_{\Delta E}$ around the fitted $\Delta E$ peak, where
$\sigma_{\Delta E}$ is the $\Delta E$ resolution and
$E_{\bar{D}^0}$ is the reconstructed $\bar{D}^0$
energy in the  initial $e^+e^-$ rest frame. The explicit $\Delta E$ requirements for the three ST modes are
listed in Table~\ref{tab:numST}.

\begin{figure}[tp!]
\begin{center}
\includegraphics[width=\linewidth]{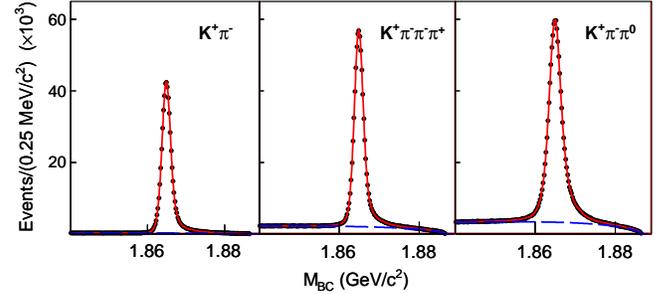}
\caption{(Color online)~The $M_{\rm BC}$ distributions for the three ST modes. The points are data, the (red) solid curves are the projection of the sum of all fit components and the (blue) dashed curves are the projection of the background component of the fit.}
\label{fig:tag_md0}
\end{center}
\end{figure}

The distributions of the variable $M_{\rm BC}$ for the three ST
modes are shown in Fig.~\ref{fig:tag_md0}. Maximum likelihood fits to the $M_{\rm BC}$ distributions are performed. The signal shape is derived from the convolution of the MC-simulated signal template function with a double-Gaussian function to account for resolution difference between MC simulation and data. An ARGUS function~\cite{plb241_278} is used to describe the combinatorial background shape.
For each tag mode, the ST yield is obtained by integrating the signal
function over the $D^0$ signal region specified in Table~\ref{tab:numST}.
In addition to the combinatorial background, there are
also small wrong-sign (WS) peaking backgrounds in the ST $\bar{D}^0$ samples, which are from 
the doubly Cabibbo-suppressed decays of $\bar{D}{}^0\rightarrow K^-\pi^+$, $K^-\pi^+\pi^0$ and $K^-\pi^+\pi^+\pi^-$.
The $\bar{D}{}^0\rightarrow K^0_SK^-\pi^+$, $K^0_S\rightarrow \pi^+\pi^-$ decay shares the same final states as the WS background of $\bar{D}^0\rightarrow K^-\pi^+\pi^+\pi^-$.
The sizes of these WS peaking backgrounds are estimated from simulation, and are subtracted from the corresponding ST yields.
The background-subtracted ST yields are listed in Table~\ref{tab:numST}. The total ST yield summed over all three ST modes is $N_{\rm ST}=(2277.2\pm2.3)\times10^{3}$, where the uncertainty is statistical only.

\begin{table}[tp!]
\caption{ The selection requirements on $\Delta E$, the signal region in the $M_{\rm BC}$ distribution, and the background-subtracted ST yields $N_{\rm ST}$ in data for each of the three tag decay modes. }
\begin{center}
\begin{tabular}
{lccc} \hline\hline Decay      & $\Delta E$ (GeV)        & Signal Region     &  $N_{\rm ST}$ ($\times 10^3$)      \\
Mode & & (GeV/$c^2$) &  \\
\hline $K^+\pi^-$                & [$-$0.025, 0.028]       & [$1.860, 1.875$]               &  ~$540.2~\pm0.8~$    \\
       $K^+\pi^-\pi^-\pi^+$      & [$-$0.020, 0.023]       & [$1.860, 1.875$]               &  ~$701.1~\pm1.7~$  \\
       $K^+\pi^-\pi^0$           & [$-$0.044, 0.066]       & [$1.858, 1.875$]               &  $1035.9~\pm1.3~$  \\
\hline\hline
\end{tabular}
\label{tab:numST}
\end{center}
\end{table}

\begin{linenumbers}
Candidates for the SL decay $D^0\rightarrow \bar{K}^0\pi^-e^+\nu_e$ are selected from the remaining tracks recoiling against the ST $\bar{D}^0$ mesons. The $\bar{K}^0$ meson is reconstructed as a $K^0_S$. The $K^0_S$ mesons are reconstructed from two oppositely charged tracks and the invariant mass of the $K^0_S$ candidate is required to be within $(0.485,~0.510)$~GeV$/c^2$. For each $K_S^0$ candidate, a fit is applied to constrain the two charged tracks to a common vertex, and this $K^0_S$ decay vertex is required to be separated from the interaction point by more than twice the standard deviation of the measured flight distance. A further requirement is that there must only be two other tracks in the event and that they must be of opposite charge. The electron hypothesis is assigned to the track that has the same charge as that of the kaon on the tag side. For electron particle identification~(PID), the specific
ionization energy losses measured by the MDC, the time of flight, and the shower properties from the electromagnetic
calorimeter (EMC) are used to construct likelihoods for electron, pion and kaon hypotheses ($\mathcal{L}_e$,
$\mathcal{L}_\pi$ and $\mathcal{L}_K$).  The electron candidate must satisfy $\mathcal{L}_{e} > 0.001$ and
$\mathcal{L}_e/(\mathcal{L}_e+\mathcal{L}_{\pi}+\mathcal{L}_K)>0.8$. Additionally, the EMC energy of the
electron candidate has to be more than 70\% of the track momentum measured in the MDC~($E/p>0.7c$).
The energy loss due to bremsstrahlung is partially recovered by adding the energy of the
EMC showers that are within 5$^{\circ}$ of the electron direction and not matched to other
particles~\cite{bes3electronSL}. The pion hyphotesis is assigned to the remaining charged track and must satisfy the same criteria as in Ref.~\cite{prl118_112001}. The background from $D^0\rightarrow \bar{K}^0\pi^+\pi^-$ decays reconstructed as $D^0\rightarrow \bar{K}^0\pi^-e^+\nu_e$ is rejected by requiring the $\bar{K}^0\pi^-e^+$ invariant mass ($M_{\bar K^0\pi^-e^+}$) to be less than 1.80~GeV/$c^2$. The backgrounds associated with fake photons are suppressed by requiring the maximum energy of any unused photon ($E_{\gamma\,{\rm max}}$) to be less than 0.25~GeV.

The energy and momentum carried by the neutrino are denoted by $E_{\rm miss}$ and $\vec{p}_{\rm miss}$, respectively. They are calculated from
the energies and momenta of the tag ($E_{\bar{D}^0}$, $\vec{p}_{\bar{D}^0}$) and the measured SL decay products ($E_{\rm SL}=E_{\bar{K}^0}+E_{\pi^-}+E_{e^+}$,
$\vec{p}_{\rm SL}=\vec{p}_{\bar{K}^0}+\vec{p}_{\pi^-}+\vec{p}_{e^+}$) using the relations $E_{\rm miss}=\sqrt{s}/2-E_{\rm SL}$ and $\vec{p}_{\rm miss}=\vec{p}_{D^0}-\vec{p}_{\rm SL}$   in the initial $e^+e^-$ rest frame. 
Here, the momentum $\vec{p}_{D^0}$ is given by
$\vec{p}_{D^0}=-\hat{p}_{\rm tag}\sqrt{(\sqrt{s}/2)^2-m^2_{\bar{D}^0}},$
where $\hat{p}_{\rm tag}$ is the
momentum direction of the ST $\bar{D}^0$ and $m_{\bar{D}^0}$ is the nominal $\bar{D}^0$ mass~\cite{pdg16}. Information on the undetected neutrino is obtained by using the variable $U_{\rm miss}$ defined by
\begin{equation}
U_{\rm miss} \equiv E_{\rm miss}-|\vec{p}_{\rm miss}| .
\end{equation}
The $U_{\rm miss}$ distribution
is expected to peak at zero for signal events.
\end{linenumbers}

Figure~\ref{fig:formfactor}(a) shows the $U_{\rm miss}$ distribution of the accepted candidate events for $D^0\rightarrow \bar{K}^{0}\pi^-e^+\nu_e$ in data.
To obtain the signal yield, an unbinned maximum likelihood fit of the $U_{\rm miss}$ distribution is performed.
In the fit, the signal is described with a shape derived from the simulated signal events convolved with a Gaussian function, where the width of the Gaussian function is determined by the fit. The
background is described by using the shape obtained from the MC simulation.
The yield of DT $D^0\rightarrow \bar{K}^0\pi^-e^+\nu_e$ events is determined to be $3131\pm64({\rm stat.})$.
The backgrounds from the non-$D^0$ and non-$K_S^0$ decays are estimated by examining
the ST candidates in the $M_{\rm BC}$ sideband, defined in the range $(1.830, 1.855)$~GeV/$c^2$, and the SL candidates in the $K^0_S$ sidebands, defined in the ranges $(0.450, 0.475)$~GeV/$c^2$ or $(0.525, 0.550)$~GeV/$c^2$ in data, respectively. The yield of this type of background is estimated to be $19.4\pm5.3$.
After subtracting these background events, we evaluate the number of the signal DT events to be $N_{\rm DT}=3112\pm64({\rm stat.})$.

The detection efficiency $\varepsilon_{\rm SL}$ is estimated to be $(9.53\pm0.01)\%$, and the BF of $D^0\rightarrow \bar{K}^{0}\pi^-e^+\nu_e$ is determined as $\mathcal B({D^0\rightarrow \bar{K}^{0}\pi^-e^+\nu_e})=(1.434\pm0.029({\rm stat.}))\%$. Due to the double tag technique, the BF measurement is insensitive to the systematic uncertainty in the ST efficiency.
The uncertainties due to the pion and electron tracking efficiencies are estimated to be 0.5\%~\cite{prd92_072012} and the uncertainties due to their PID efficiencies are estimated to be 0.5\%~\cite{prd92_072012}, where the tracking and PID uncertainties are conservatively estimated to account for the possible differences of the momentum spectra in Ref.~\cite{prd92_072012}.
The uncertainty due to the $\bar{K}^0$ reconstruction is 1.5\%~\cite{prl118_112001}. The uncertainty due to the $E/p$ requirement is 0.4\%~\cite{prd94_032001}.
The uncertainty associated with the $E_{\gamma\,{\rm \max}}$ requirement is estimated to be 0.4\% by analyzing the DT $D^0\bar{D}^0$ events where both $D$ mesons decay to hadronic final states.
The uncertainty due to the modeling of the signal in simulated events is estimated to be 0.8\% by varying
the input form factor parameters by $\pm 1\sigma$ as determined in this work. The uncertainty associated with the fit of the $U_{\rm miss}$ distribution is estimated to be 0.7\% by varying the fitting ranges and the shapes which parametrize the signal and background. The uncertainty associated with the fit of the $M_{\rm BC}$ distributions used to determine $N_{\rm ST}$ is 0.5\% and is evaluated by varying the bin size, fit range and background distributions. Further systematic uncertainties are assigned due to the statistical precision of the simulation (0.2\%), the background subtraction (0.2\%), and the input BF of the decay $K^0_S\rightarrow \pi^+ \pi^-$ (0.1\%). The systematic uncertainty contributions are summed in quadrature, and the total systematic uncertainty on the BF measurement is 2.2\% of the central value.

\section{$D^0\rightarrow \bar{K}^{0}\pi^- e^+\nu_{e}$ Decay rate formalism}
The differential decay width of $D^0\rightarrow \bar{K}^{0}\pi^- e^+\nu_{e}$ can be expressed in terms of five kinematic variables:
the square of the invariant mass of the $\bar{K}^0\pi^-$ system $m_{\bar{K}^0\pi^-}^2$, the square of the invariant mass of the $e^+\nu_e$ system ($q^2$), the angle between the $\bar{K}^0$ and the $D^0$ direction in the $\bar{K}^0\pi^-$ rest frame ($\theta_{\bar{K}^0}$), the angle between the $\nu_{e}$ and the $D^0$ direction in the $e^+\nu_e$ rest frame ($\theta_e$), and the acoplanarity angle between the two decay planes ($\chi$).
Neglecting the mass of $e^+$, the differential decay width of $D^0\rightarrow \bar{K}^{0}\pi^- e^+\nu_{e}$ can be expressed as~\cite{prd46_5040}
\begin{eqnarray}
d^5\Gamma&=&\frac{G^2_F|V_{cs}|^2}{(4\pi)^6m^3_{D^0}}X\beta \mathcal{I}(m_{\bar{K}^0\pi^-}^2, q^2, \theta_{\bar{K}^0}, \theta_e, \chi) \nonumber \\
         && dm_{\bar{K}^0\pi^-}^2dq^2d{\rm cos}\theta_{\bar{K}^0}d{\rm cos}\theta_ed\chi,
\end{eqnarray}
where $X=p_{\bar{K}^{0}\pi^-}m_{D^0}$, $\beta=2p^{*}/m_{\bar{K}^{0}\pi^-}$, and $p_{\bar{K}^{0}\pi^-}$ is the momentum of the $\bar{K}^{0}\pi^-$ system in the rest $D^0$ system and $p^*$ is the momentum of $\bar{K}^{0}$ in the $\bar{K}^{0}\pi^-$ rest frame. The Fermi coupling constant is denoted by $G_F$.
The dependence of the decay density $\mathcal{I}$ is given by
\begin{eqnarray}
\mathcal{I}&=&\mathcal{I}_1+\mathcal{I}_2{\rm cos2}\theta_e+\mathcal{I}_3{\rm sin}^2\theta_e{\rm cos}2\chi+\mathcal{I}_4{\rm sin}2\theta_e{\rm cos}\chi  \nonumber\\
           &+&\mathcal{I}_5{\rm sin}\theta_e{\rm cos}\chi+\mathcal{I}_6{\rm cos}\theta_e+\mathcal{I}_7{\rm sin}\theta_e{\rm sin}\chi \nonumber \\
           &+&\mathcal{I}_8{\rm sin}2\theta_e{\rm sin}\chi+\mathcal{I}_9{\rm sin}^2\theta_e{\rm sin}2\chi,
\label{eq:Ifunc}
\end{eqnarray}
where $\mathcal{I}_{1,...,9}$ depend on $m_{\bar{K}^{0}\pi^-}^2$, $q^2$ and $\theta_{\bar{K}^0}$~\cite{prd46_5040} and can be expressed in terms of three form factors, $\mathcal{F}_{1,2,3}$. The form factors can be expanded into partial waves including $S$-wave ($\mathcal{F}_{10}$), $P$-wave ($\mathcal{F}_{i1}$) and $D$-wave ($\mathcal{F}_{i2}$), to show their explicit dependences on $\theta_{\bar{K}^0}$.
Analyses of the decay $D^+\rightarrow K^+\pi^-e^+\nu_e$ by using much higher statistics performed by the BABAR~\cite{prd83_072001} and BESIII~\cite{prd94_032001} collaborations do not observe a $D$-wave component and hence it is not considered in this analysis. Consequently, the form factors can be written as
\begin{equation}
\mathcal{F}_1=\mathcal{F}_{10}+\mathcal{F}_{11}\cos\theta_{\bar{K}^0},\mathcal{F}_2=\frac{1}{\sqrt{2}}\mathcal{F}_{21},\mathcal{F}_3=\frac{1}{\sqrt{2}}\mathcal{F}_{31},
\label{eq:F1}
\end{equation}
where $\mathcal{F}_{11}$, $\mathcal{F}_{21}$ and $\mathcal{F}_{31}$ are related to the helicity basis form factors $H_{0,\pm}(q^2)$~\cite{prd46_5040,RevModPhys67_893}.
The helicity form factors can in turn be related to the two axial-vector form factors, $A_1(q^2)$ and $A_2(q^2)$, as well as the vector form factor $V(q^2)$.
The $A_{1,2}(q^2)$ and $V(q^2)$ are all taken as the simple pole form
$A_i(q^2)=A_{1,2}(0)/(1-q^2/M^2_A)$ and $V(q^2)=V(0)/(1-q^2/M^2_V)$, with pole masses $M_V=M_{D_s^*(1^-)}=2.1121$~GeV/$c^2$~\cite{pdg16} and $M_A=M_{D_s^*(1^+)}=2.4595$~GeV/$c^2$~\cite{pdg16}.
The form factor $A_1(q^2)$ is common to all three helicity amplitudes. Therefore, it is natural to define two form factor ratios as $r_V=V(0)/A_1(0)$ and $r_2=A_2(0)/A_1(0)$
at the momentum square $q^2=0$.

The amplitude of the $P$-wave resonance $\mathcal{A}(m)$ is expressed as~\cite{prd94_032001,prd83_072001}
\begin{equation}
\mathcal{A}(m)=\frac{m_0\Gamma_0(p^*/p^*_0)}{m_0^2-m^2_{\bar{K}^{0}\pi^-}-im_0\Gamma(m_{\bar{K}^{0}\pi^-})}\frac{B(p^*)}{B(p^*_0)},
\end{equation}
where $B(p)=\frac{1}{\sqrt{1+R^2p^2}}$ with $R=3.07$~GeV$^{-1}$~\cite{prd94_032001} and
$\Gamma \left(m_{\bar{K}^{0}\pi^-}\right)=\Gamma_0\left(\frac{p^*}{p^*_0}\right)^3\frac{m_0}{m_{\bar{K}^{0}\pi^-}}\left[\frac{B\left(p^*\right)}{B\left(p^*_0\right)}\right]^2$,
where $p^*_0$ is the momentum of $\bar{K}^0$ at the pole mass of the resonance $m_0$, and $\alpha=\sqrt{3\pi \mathcal{B}_{K^*}/(p^*_0\Gamma_0)}$, $\mathcal{B}_{K^*}=\mathcal{B}(K^{*}(892)^-\rightarrow \bar{K}^0\pi^-)$.

The $S$-wave related $\mathcal{F}_{10}$ is described by~\cite{prd94_032001,prd83_072001}
\begin{equation}
\mathcal{F}_{10}=p_{\bar{K}^{0}\pi^-}m_{D^0}\frac{1}{1-\frac{q^2}{m^2_A}}\mathcal{A}_S(m),
\end{equation}
where the term $\mathcal{A}_S(m)$ corresponds to the mass-dependent $S$-wave amplitude, and the same expression of $\mathcal{A}_S(m)=r_SP(m)e^{i\delta_S(m)}$ as in Refs.~\cite{prd94_032001,prd83_072001} is adopted, in which $P(m)=1+xr_S^{(1)}$ with $x=\sqrt{\left(\frac{m}{m_{\bar{K}^0}+m_{\pi^-}}\right)^2-1}$,
and $\delta_S(m)=\delta^{1/2}_{\rm BG}$ with $\cot(\delta^{1/2}_{\rm BG})=1/(a^{1/2}_{\rm S,BG}p^*)+b^{1/2}_{\rm S,BG}p^*/2$.

\begin{figure}[tp!]
\begin{center}
   \flushleft
   \begin{minipage}[t]{8.8cm}
   \includegraphics[width=\linewidth]{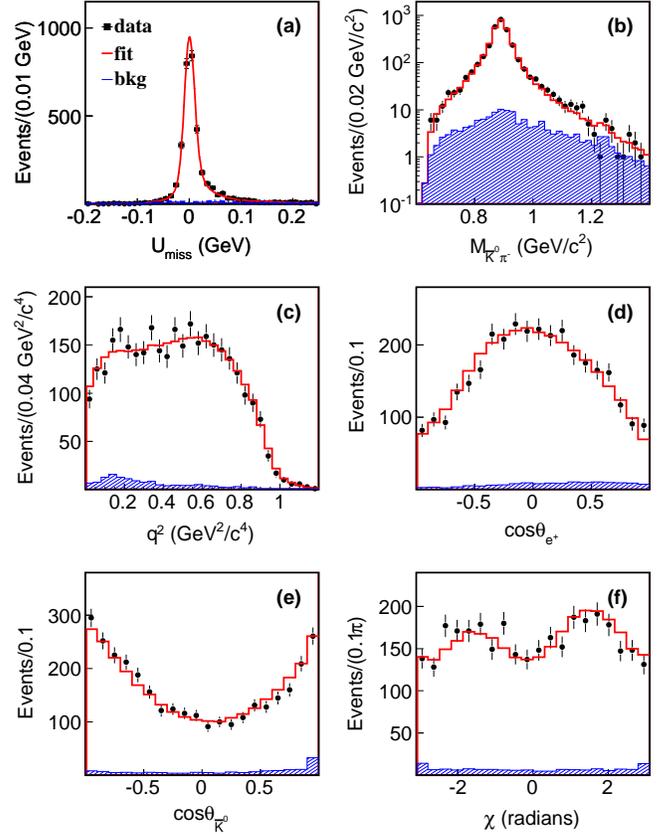}
   \end{minipage}
   \caption{ (Color online)~(a) Fit to $U_{\rm miss}$ distribution of the SL candidate events.
   Projections onto five kinematic variables (b) $M_{\bar{K}^0\pi^-}$, (c) $q^2$, (d) $\cos\theta_{e^+}$, (e) $\cos\theta_{\bar{K}^0}$, and (f) $\chi$ for $D^0\rightarrow \bar{K}^0\pi^-e^+\nu_e$. The dots with error bars are data, the red curve/histograms are the fit results, and the shadowed histograms are the simulated background.}
\label{fig:formfactor}
\end{center}
\end{figure}

An unbinned five-dimensional maximum likelihood fit to the distributions of $m_{\bar{K}^0\pi^-}$, $q^2$, $\cos\theta_{e^+}$, $\cos\theta_{\bar{K}^0}$, and $\chi$ for the
$D^0\rightarrow \bar{K}^{0}\pi^- e^+\nu_{e}$ events within $-0.10<U_{\rm miss}<0.15$~GeV is performed in a similar manner to Ref.~\cite{prd94_032001}.
The projected distributions of the fit onto the fitted variables are shown in
Figs.~\ref{fig:formfactor}\,(b-f).
In this fit, the parameters of $r_V$, $r_2$, $m_0$, $\Gamma_0$, $r_S$ and $a^{1/2}_{\rm S,BG}$ are float,
while $r_S^{(1)}$ and $b^{1/2}_{\rm S,BG}$ are fixed to $0.08$ and $-0.81$~(GeV/$c$)$^{-1}$ due to limited statistics, respectively, based on the analysis of $D^+\rightarrow K^+\pi^-e^+\nu_e$ at BESIII~\cite{prd94_032001}.
The fit results are summarized in Table~\ref{tab:FitResults}.
The goodness of fit is estimated by using the $\chi^2/{\rm ndof}$, where ${\rm ndof}$ denotes the number of degrees of freedom. The $\chi^2$ is
calculated from the comparison between the measured and expected number of events in the five-dimensional space of the kinematic variables $m_{\bar{K}^0\pi^-}$, $q^2$, $\cos\theta_{e^{+}}$, $\cos\theta_{\bar{K}^0}$, and $\chi$ which are initially divided into 2, 2, 3, 3, and 3 bins, respectively.
The bins are set with different sizes, so that they contain sufficient numbers of signal events for credible $\chi^2$ calculation. Each five-dimensional bin is required to contain at least ten events; otherwise, it is combined with an adjacent bin. The $\chi^2$ value is
calculated as
\begin{equation}
\chi^2=\displaystyle{\sum_i^{\rm N_{\rm bin}}\frac{(n_i^{\rm data}-n_i^{\rm fit})^2}{n_i^{\rm fit}}},
\end{equation}
where $N_{\rm bin}$ is the number of bins, $n_i^{\rm data}$ denotes the
measured number of events of the $i$-th bin, and $n_i^{\rm fit}$ denotes the the expected number of events of the $i$th bin. The ${\rm ndof}$ is the number of bins minus
the number of fit parameters minus 1. The $\chi^2/{\rm ndof}$ obtained is 96.3/98, which shows a good fit quality. The fit procedure is validated using a large simulated sample of inclusive events, where the pull distribution of each fitted parameter is found to be consistent with a normal distribution.

\begin{table}
\begin{center}
\caption{The fit results, where the first uncertainties are statistical and the second are systematic. } \normalsize
\begin{tabular}
{ll} \hline\hline  Variable~~~~~~~~~~~~~~~~~~~~~~~     &  ~~~~~~~~~~~~~~~Value                       \\ \hline
$M_{K^{*}(892)^-}$ (MeV/$c^2$)                                      & ~~~$891.7~\pm0.6~\pm0.2$          \\
$\Gamma_{K^{*}(892)^-}$ (MeV)                                       & ~~~~~$48.4~\pm1.5~\pm0.5$          \\
$r_S$ (GeV)$^{-1}$                                                  & ~$-11.21\pm1.03\pm1.15$       \\
$a^{1/2}_{\rm S,BG}$ (GeV/$c$)$^{-1}$                               &~~~~~~$1.58\pm0.22\pm0.18$     \\
$r_V$                                                               &~~~~~~$1.46\pm0.07\pm0.02$     \\
$r_2$                                                               &~~~~~~$0.67\pm0.06\pm0.01$     \\
\hline\hline
\end{tabular}
\label{tab:FitResults}
\end{center}
\end{table}

\begin{table*}
\begin{center}
\caption{Systematic uncertainties (in \%) of the fitted parameters. }
\begin{tabular}
{lcccccccccc} \hline\hline \normalsize
Parameter~~~~~& ~$E_{\gamma\,{\rm \max}}$~ &  ~~$E/p$~~ & ~~~~$f$~~~~ & ~${\rm Tracking\&PID}$~ & ~$D$-wave~ & ~~$r_S^{(1)}$~~ & ~$b^{1/2}_{\rm S,BG}$~ & ~~${\rm Total}$~~ \\ \hline
$M_{K^{*}(892)^-}$    &   0.00 & 0.01 & 0.00 & 0.00 & 0.01 & 0.01 & 0.01 &~~0.02  \\
$\Gamma_{K^{*}(892)^-}$ & 0.52 & 0.95 & 0.23 & 0.04 & 0.12 & 0.08 & 0.12 &~~1.12  \\
$r_S$                 &   4.45 & 1.85 & 2.58 & 0.24 & 0.76 & 8.57 & 1.26 & 10.27  \\
$a^{1/2}_{\rm S,BG}$  &   7.66 & 3.52 & 1.36 & 0.26 & 0.87 & 0.11 & 7.78 & 11.59  \\
$r_V$                 &   0.34 & 0.83 & 0.37 & 0.57 & 0.12 & 0.29 & 0.42 &~~1.21  \\
$r_2$                 &   0.95 & 0.27 & 0.30 & 0.02 & 0.27 & 0.03 & 0.60 &~~1.22  \\
$f_{K^{*}(892)^-}$    &   0.52 & 0.22 & 0.28 & 0.03 & 0.10 & 0.07 & 0.16 &~~0.66  \\
$f_{S-{\rm wave}}$      &   8.89 & 3.81 & 4.72 & 0.54 & 1.81 & 1.09 & 2.54 & 11.27  \\
\hline\hline
\end{tabular}
\label{tab:Syserr}
\end{center}
\end{table*}

The fit fraction of each component can be determined by the ratio of the decay intensity of the specific component
and that of the total. The fractions of $S$-wave and $P$-wave ($K^{*}(892)^-$) are found to be
$f_{S-{\rm wave}}=(5.51\pm0.97({\rm stat.}))\%$ and $f_{K^{*}(892)^-}=(94.52\pm0.97({\rm stat.}))\%$, respectively.

The systematic uncertainties of the fitted parameters and the fractions of $S$-wave and $K^{*}(892)^-$ components
are defined as the difference between the fit results in nominal conditions and those obtained after
changing a variable or a condition by an amount which corresponds to an estimate of the uncertainty in the determination
of this quantity.
The systematic uncertainties due to the $E_{\gamma\,{\rm \max}}$ and $E/p$ requirements are estimated by using alternative requirements of
$E_{\gamma\,{\rm \max}}<0.20$~GeV and $E/p>0.75$, respectively. The systematic uncertainty because of the background fraction ($f$) is estimated by varying its value by $\pm 10\%$ which is the difference of the background fractions in the selected ST $\Delta E$ regions between data and MC simulation.
The systematic uncertainties arising from the requirements placed on the charged pion, the electron and the $K^0_S$ are estimated by varying the pion/electron tracking and PID efficiencies, and $K_S^0$ detection efficiency by $\pm0.5\%$, $\pm0.5\%$ and $\pm1.5\%$, respectively.
The systematic uncertainty due to neglecting a possible contribution from the $D$-wave component is estimated by incorporating the $D$-wave component in Eq.~(\ref{eq:F1}). The systematic uncertainties in the fixed parameters of $r_S^{(1)}$ and $b^{1/2}_{\rm S,BG}$ are estimated
by varying their nominal values by $\pm1\sigma$. 
All of the variations mentioned above will result in differences of the fitted parameters and the extracted fractions of $S$-wave and $K^{*}(892)^-$ components from that under the nominal conditions. 
These differences are assigned as the systematic uncertainties and 
summarized in Table III, where the total systematic uncertainty is obtained by adding all contributions in quadrature.

\section{Summary}
In summary, using $2.93~\mathrm{fb}^{-1}$ of data collected at $\sqrt{s}=3.773$ GeV by the BESIII detector, the absolute BF of $D^0\rightarrow \bar{K}^0\pi^-e^+\nu_{e}$ is measured to be $\mathcal{B}(D^0\rightarrow \bar{K}^0\pi^-e^+\nu_{e})=(1.434\pm0.029({\rm stat.})\pm0.032({\rm syst.}))\%$, which is significantly more precise than the current world-average value~\cite{pdg16}.
The first analysis of the dynamics of $D^0\rightarrow \bar{K}^0\pi^-e^+\nu_{e}$ decay is performed and the $S$-wave component is observed with a fraction $f_{S-{\rm wave}}=(5.51\pm0.97({\rm stat.})\pm0.62({\rm syst.}))\%$, leading to $\mathcal{B}[D^0\rightarrow (\bar{K}^0\pi^-)_{S-{\rm wave}}e^+\nu_e]=(7.90\pm1.40({\rm stat.})\pm0.91({\rm syst.}))\times 10^{-4}$.

The $P$-wave component is observed with a fraction of $f_{K^{*}(892)^-}=(94.52\pm0.97({\rm stat.})\pm0.62({\rm syst.}))\%$ and the corresponding BF is given as $\mathcal{B}(D^0\rightarrow K^{*-}e^+\nu_e)=(2.033\pm0.046({\rm stat.})\pm0.047({\rm syst.}))\%$. It is consistent with, and more precise than, the result from the CLEO collaboration~\cite{prl95_181802}.
In addition, the form factor ratios of the $D^0\rightarrow K^{*}(892)^-e^+\nu_{e}$ decay are determined to be
$r_V=1.46\pm0.07({\rm stat.})\pm0.02({\rm syst.})$ and
$r_2=0.67\pm0.06({\rm stat.})\pm0.01({\rm syst.})$.
They are consistent with the measurements from the FOCUS collaboration~\cite{plb607_67} using the decay $D^0\rightarrow \bar{K}^0\pi^-\mu^+\nu_{\mu}$ within uncertainties, but with significantly improved precision. 

The BESIII collaboration thanks the staff of BEPCII and the IHEP computing center for their strong support. This work is supported in part by National Key Basic Research Program of China under Contract No. 2015CB856700; National Natural Science Foundation of China (NSFC) under Contracts Nos. 11335008, 11425524, 11505010, 11625523, 11635010, 11735014, 11775027; the Chinese Academy of Sciences (CAS) Large-Scale Scientific Facility Program; the CAS Center for Excellence in Particle Physics (CCEPP); Joint Large-Scale Scientific Facility Funds of the NSFC and CAS under Contracts Nos. U1532257, U1532258, U1732263; CAS Key Research Program of Frontier Sciences under Contracts Nos. QYZDJ-SSW-SLH003, QYZDJ-SSW-SLH040; 100 Talents Program of CAS; INPAC and Shanghai Key Laboratory for Particle Physics and Cosmology; German Research Foundation DFG under Contracts Nos. Collaborative Research Center CRC 1044, FOR 2359; Istituto Nazionale di Fisica Nucleare, Italy; Koninklijke Nederlandse Akademie van Wetenschappen (KNAW) under Contract No. 530-4CDP03; Ministry of Development of Turkey under Contract No. DPT2006K-120470; National Science and Technology fund; The Swedish Research Council; U. S. Department of Energy under Contracts Nos. DE-FG02-05ER41374, DE-SC-0010118, DE-SC-0010504, DE-SC-0012069; University of Groningen (RuG) and the Helmholtzzentrum fuer Schwerionenforschung GmbH (GSI), Darmstadt. This paper is also supported by Beijing municipal government under Contract Nos. KM201610017009, 2015000020124G064, CIT\&TCD201704047.



\begin{thebibliography}{99}
\bibitem{chargeneutral} Throughout this paper, mesons with no explicit charge denote either a charged or neutral meson;
the charge-conjugate modes are implied, unless explicitly noted.
\end{thebibliography}
\end{document}